\documentclass{emulateapj}
\submitted{ApJ, in press}
\newcommand{\kms}{\,km\,s$^{-1}$}
\newcommand{\ergs}{\,erg\,s$^{-1}$}

\newcommand{\ppdot}{\mbox{$P$-$\dot{P}$}}
\newcommand{\pdot}{\mbox{$\dot{P}$}}

\newcommand{\cxo}{\emph{Chandra}}
\newcommand{\xmm}{\emph{XMM-Newton}} 
\newcommand{\nh}{$N_{\rm H}$}

\begin{document}
\title{HUNTING FOR ORPHANED CENTRAL COMPACT OBJECTS AMONG RADIO PULSARS}
\author{J. Luo\altaffilmark{1}, C.-Y. Ng\altaffilmark{1},
W. C. G. Ho\altaffilmark{2},
S.~Bogdanov\altaffilmark{3},
V. M. Kaspi\altaffilmark{4},
and
C. He\altaffilmark{5}}
\altaffiltext{1}{Department of Physics, the University of Hong Kong, Hong
Kong; \url{ncy@bohr.physics.hku.hk}}
\altaffiltext{2}{Mathematical Sciences and STAG Research Centre, University of
Southampton, Southampton SO17 1BJ, UK}
\altaffiltext{3}{Columbia Astrophysics Laboratory, Columbia University, New
York, NY 10027, USA}
\altaffiltext{4}{Department of Physics and McGill Space Institute, McGill
University, Montreal, QC H3A 2T8, Canada}
\altaffiltext{5}{Department of Physics, the University of Chicago, Chicago, IL
60637, USA}

\shorttitle{Hunting for CCOs}
\shortauthors{Luo et al.}


\begin{abstract}
Central compact objects (CCOs) are a handful of young neutron stars found at
the center of supernova remnants (SNRs). They show high thermal X-ray
luminosities but no radio emission. Spin-down rate measurements of the three
CCOs with X-ray pulsations indicate surface dipole fields much weaker than
those of typical young pulsars. To investigate if CCOs and known radio pulsars
are objects at different evolutionary stages, we carried out a census of all
weak-field ($<10^{11}$\,G) isolated radio pulsars in the Galactic plane to
search for CCO-like X-ray emission. None of the 12 candidates are detected at
X-ray energies, with luminosity limits of $10^{32}-10^{34}$\ergs. We consider
a scenario in which the weak surface fields of CCOs are due to a rapid accretion
of supernova materials and show that as the buried field diffuses back to the
surface, a CCO descendant is expected to leave the \ppdot\ parameter space of
our candidates at a young age of a few $\times$10\,kyr. Hence, the candidates
are likely to just be old ordinary pulsars in this case. We suggest that
further searches for orphaned CCOs, which are aged CCOs with parent SNRs 
that have dissipated, should include pulsars with stronger magnetic fields.
\end{abstract}

\section{Introduction}
Before the mid-1990s it was believed that young neutron stars are all
fast-spinning objects with high surface magnetic field strengths of $\sim
10^{12}$\,G, emitting radio pulses. However, recent discoveries of new
populations of neutron stars, including central compact objects (CCOs),
magnetars, and X-ray dim isolated neutron stars, have challenged this simple
picture \citep[see reviews by][]{kas10,har13}. CCOs are the most enigmatic
class. They are found at the center of supernova remnants (SNRs) and cannot be
as easily classified as other types of objects. There are nine confirmed CCOs
sharing the following properties: (1) they are located near the centers of
young SNRs; (2) they show no radio or optical counterparts; (3) they have no
detectable pulsar wind nebulae; and (4) they exhibit a thermal spectrum in the
soft X-ray band with high luminosity $\gtrsim10^{33}$\ergs\ \citep[see][for
reviews]{del08,gh08,gha13}. Since CCOs are generally associated with very
young SNRs, their nature and evolution are highly relevant to the neutron star
production rate and the physics underlying the branching ratios of core
collapse \citep{kk08}. However, the active lifetime and evolution of CCOs are
poorly understood due to the small sample. It is also unclear if these neutron
stars are intrinsically radio-quiet, or if they are radio pulsars beamed away
from us \citep[see e.g.,][]{ho13b}.

\begin{figure}[!ht]
\epsscale{1.1}
\plotone{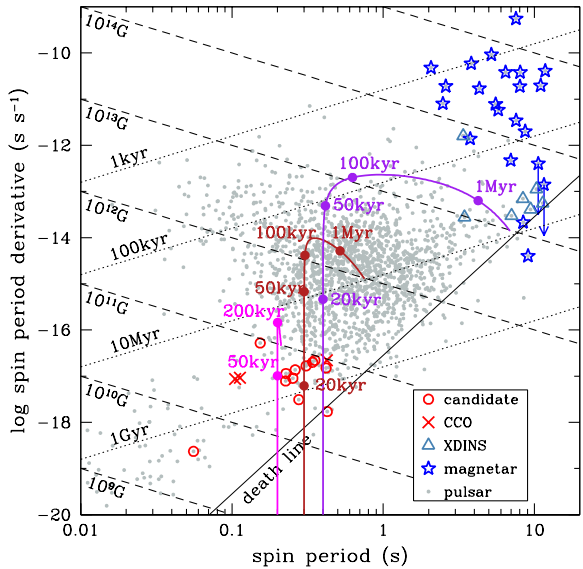}
\caption{Pulsar \ppdot\ diagram showing our orphaned CCO candidates and
other types of neutron stars. The solid curves show the theoretical evolution
of stars with buried magnetic fields at birth (see Section~\ref{sect:discuss}
for details). From left to right, the initial surface $B$-field strengths are
$10^{12}$\,G (pink), $10^{13}$\,G (brown), and $10^{14}$\,G (purple), and the
initial spin periods are set to 0.2, 0.3, and 0.4\,s, respectively, for
clarity. Selected time points are marked.
\label{fig:ppdot}}
\end{figure}

Only three CCOs have X-ray pulsations firmly detected. They show periods of
$P=0.1$--0.4\,s and long-term timing revealed small period derivatives of
$\dot P\approx10^{-17}$ \citep{hg10,gha13}. Their spin parameters are plotted
in the \ppdot\ diagram in Figure~\ref{fig:ppdot}. These suggest spin-down
luminosities of $\dot E \equiv 4 \pi^2 I\dot P / P^3 =
10^{31}$--$10^{32}$\ergs, where $I$ is the neutron star moment of inertia.
These values are nearly two orders of magnitude smaller than the CCO's X-ray
luminosities, implying that the sources cannot be entirely powered by
rotation. The characteristic age $\tau_c=P/2\dot P$ of the three CCOs is over
$10^8$\,years, much older than their true ages of a few thousand years,
estimated from their associated SNRs. The inferred surface dipole field
strengths $B\equiv3.2\times10^{19} (P\dot P)^{1/2}$\,G are of the order of
$10^{10}$\,G. These are much lower than those of young pulsars but still lie
well above the radio pulsar ``death line'' \citep[e.g., $B/P^2\simeq 1.7\times
10^{11}$\,G\,s$^{-2}$;][]{bwh+92}, which is the (uncertain) theoretical limit
for producing radio emission. Indeed, radio pulsars have been detected in the
CCO range of $P$ and \pdot\ (see Figure~\ref{fig:ppdot}), hence with
comparable dipole field strengths. This raises a fundamental question: are
CCOs and radio pulsars the same class of objects or do they belong to
disjointed sets of neutron stars?

If CCOs are ordinary radio pulsars born with weak magnetic fields \citep[see,
e.g.,][]{bub06,spr08}, their small $\dot P$ values imply very slow spin
evolution, such that they take a long time to reach the death line. The
detection of nine young CCOs implies that there should be over $10^6$ older
ones in the Galaxy \citep{kas10}. This is in contrast to the small number of
known radio pulsars with low $B\sim10^{10}$\,G as shown in the \ppdot\ diagram
in Figure~\ref{fig:ppdot}, given that there is no observational bias against
these pulsars in radio surveys \citep[see][]{kbm+03,fk06}. The discrepancy
could be reconciled if CCOs are strong-field objects but appear to have weak
surface fields at a young age. This is supported by modeling of CCOs' X-ray
light curves, which suggest a strong crustal field \citep{gph10,sl12,bog14}.
It was proposed that the $B$-fields of CCOs could be buried by supernova
fallback \citep{hg10,ho11,ho13a}, and hence the radio emission is suppressed
(as in the case of accreting millisecond pulsars in quiescence; see e.g.,
\citealt{sah+14,abp+15}). After accretion stops, the $B$-field is expected to
diffuse back to the surface with a timescale depending sensitively on the
amount of accreted mass and ranging from $10^3$\,years to over $10^6$\,years
\citep[see][]{che89,gpz99,ho11,vp12}. The radio emission would then presumably
switch on \citep{mp96}. This picture predicts that young CCOs should be
radio-quiet and aged ones could become ordinary
radio pulsars \citep[see][]{gha+13,bnk14}.

No radio emission has yet been observed from any CCO \citep[see][and
references therein]{pst04}, but given the small number of known pulsating CCOs
and the possibility of small beaming fractions, the result is inconclusive and
a more systematic study is needed. In addition to deeper radio observations,
we can turn the search around to look for CCO-like X-ray emission from
selected weak-field radio pulsars. Indeed, if CCOs manifest as radio pulsars,
this could be a more efficient way to detect them, especially after their
natal SNRs fade away in $\sim10^5$\,years. Any detection of X-ray emission
from the radio-selected sample will confirm CCOs as a subset of radio pulsars,
providing direct evidence to rule out the scenario of ongoing accretion. There
were previous attempts to identify CCOs from radio pulsars positionally
coincident with SNRs \citep{bnk14} and aged CCOs with parent SNRs that have
dissipated (so-called orphaned CCOs) from disrupted recycled pulsars
\citep{gha+13}. However, no new CCOs have been found. To complete the study,
here we present an X-ray census of all isolated weak-field radio pulsars near
the Galactic plane to search for CCO-like emission. The sample selection and
data analysis are described in Section~\ref{sect:data} and we report the
detection limits and discuss the implications in Section~\ref{sect:discuss}.

%


\section{OBSERVATIONS AND DATA REDUCTION}
\label{sect:data}
We select weak-field radio pulsars from the ATNF catalog \citep{mht+05} with
similar spin parameters as those of the known CCOs, according to the following
criteria:
\begin{enumerate}
\item weak surface dipole $B$-fields of $B\leq10^{11}$\,G inferred
from spin-down;
\item isolated and having periods $P\geq0.05$\,s to avoid recycled pulsars;
and
\item located $<100$\,pc from the Galactic plane to exclude old objects.
\end{enumerate} 
The pulsar height from the plane is calculated from the source's Galactic
latitude and estimated distance. Neutron stars are born at an average distance
of 50\,pc from the plane with a mean space velocity of $\sim350$\kms\
\citep{fk06}. At this velocity, a pulsar travels only 36\,pc in $10^5$\,years,
which is well below the cut even in the rare case that a pulsar moves exactly
perpendicular to the plane.

\begin{deluxetable*}{cccccccccc}
\tabletypesize{\small}
\tablewidth{0pt}
\tablecaption{Properties of Orphaned CCO Candidates\label{tab:candid}}
\tablehead{\colhead{Pulsar} & \colhead{Dist.\tablenotemark{a}} & \colhead{$l$}
& \colhead{$b$} & \colhead{DM} & \colhead{$N_{\mathrm{H}}$\tablenotemark{b}} &
\colhead{$P$} & \colhead{$\dot{P}$}&\colhead{$\tau_c$\tablenotemark{c}} &
\colhead{$B$\tablenotemark{d}} \\ 
& \colhead{(kpc)} & \colhead{(\arcdeg)} & \colhead{(\arcdeg)} &
\colhead{($10$\,pc\,cm$^{-3}$)} & \colhead{($10^{22}$\,cm$^{-2}$)} &
\colhead{(s)} & \colhead{$(10^{-18})$} & \colhead{($10^8$\,years)} &
\colhead{($10^{10}$\,G)} } 
\startdata
\object[PSR J0609+2130]{J0609+2130} & 1.2 & 189.2 & 1.0 & 3.9 & $0.12^{+0.09}_{-0.04}$ & 0.06 & 0.24 & 38 & 0.37 \\
\object[PSR J1107-5907]{J1107$-$5907} & 1.3 & 289.9 & 1.1 & 4.0 & $0.12^{+0.09}_{-0.04}$ & 0.25 & 9.0 & 4.5 & 4.8 \\
\object[PSR J1355-6206]{J1355$-$6206} & 8.3 & 310.3 & $-$0.2 & 55 & $1.7^{+1.2}_{-0.5}$ & 0.28 & 3.1 & 14 & 3.0 \\
\object[PSR J1425-5723]{J1425$-$5723} & 1.2 & 315.4 & 3.2 & 4.3 & $0.13^{+0.1}_{-0.04}$ & 0.35 & 22 & 2.5 & 8.9 \\
\object[PSR J1650-4341]{J1650$-$4341} & 7.5 & 341.6 & 0.5 & 67 & $2.1^{+1.5}_{-0.7}$ & 0.31 & 17 & 2.9 & 7.3 \\
\object[PSR J1653-4315]{J1653$-$4315} & 4.6 & 342.3 & 0.4 & 34 & $1.0^{+0.7}_{-0.3}$ & 0.42 & 15 & 4.4 & 8.0 \\
\object[PSR J1702-4217]{J1702$-$4217} & 7.1 & 344.1 & $-$0.3 & 63 & $1.9^{+1.4}_{-0.6}$ & 0.23 & 11 & 3.2 & 5.2 \\
\object[PSR J1739-3951]{J1739$-$3951} & 0.8 & 4.0 & $-$4.7 & 2.5 & $0.076^{+0.05}_{-0.02}$ & 0.34 & 20 & 2.7 & 8.3 \\
\object[PSR J1755-2725]{J1755$-$2725} & 2.4 & 2.4 & $-$1.1 & 12 & $0.35^{+0.25}_{-0.10}$ & 0.26 & 14 & 3.0 & 6.1 \\
\object[PSR J1810-1820]{J1810$-$1820} & 5.6 & 12.1 & 0.3 & 45 & $1.4^{+1.0}_{-0.5}$ & 0.15 & 52 & 0.47 & 9.1 \\
\object[PSR J1819-1510]{J1819$-$1510} & 5.3 & 15.9 & $-$0.1 & 42 & $1.3^{+0.9}_{-0.4}$ & 0.23 & 7.9 & 4.6 & 4.3\\
\object[PSR B1952+29]{B1952+29} & 0.7 & 65.3 & 0.8 & 0.79 & $0.024^{+0.017}_{-0.008}$ & 0.43 & 1.7 & 40 & 2.7
\enddata
\tablenotetext{a}{All distance are estimated from the DM, since no parallax
measurements are available.}
\tablenotetext{b}{\nh\ values are estimated by
$N_\mathrm{H}\rm\;(10^{20}\,cm^{-2})=
0.30^{+0.13}_{-0.09}\;DM\;(pc\,cm^{-3})$ \citep{hnk13}.} 
\tablenotetext{c}{The characteristic age is given by $\tau_c\equiv
P/2{\dot P}$.} 
\tablenotetext{d}{The surface magnetic field strength, $B$, is inferred from
the spin parameters with $B\equiv3.2\times10^{19} (P\dot P)^{1/2}$\,G, where
$P$ is in seconds.}
\end{deluxetable*} 

\begin{deluxetable*}{cccccccc}
\tabletypesize{\footnotesize}
\tablewidth{0pt}
\centering
\tablecaption{Flux and Age Limits of the Orphaned CCO Candidates
\label{tab:results}}
\tablehead{\colhead{Pulsar} & \colhead{ObsID\tablenotemark{a}} &
\colhead{Effective} & \colhead{Count Rate} & \colhead{$kT_{\rm
max}$\tablenotemark{c}} & \colhead{$L^{\rm bol}_{\rm max}$\tablenotemark{c}} &
\colhead{Age Lower}\\
& & \colhead{Exposure} & \colhead{Upper Limit\tablenotemark{b}} & & &
\colhead{Limit\tablenotemark{d}}\\
& & \colhead{(ks)} & \colhead{($10^{-4}$\,cnt\,s$^{-1}$)} & \colhead{(eV)} &
\colhead{($10^{33}$\,erg\,s$^{-1}$)} & \colhead{(kyr)}}
\startdata
J0609+2130 & \dataset[ADS/Sa.CXO#obs/12687]{12687} & 5 &6.3& 51 & $0.18_{-0.02}^{+0.04}$ & 20\\
J1107$-$5907 & \dataset[ADS/Sa.CXO#obs/12688]{12688} & 5 &14& 55 & $0.25_{-0.02}^{+0.06}$ & 20\\
J1355$-$6206 & \dataset[ADS/Sa.CXO#obs/13806]{13806} & 3.4 &9.1& 134 & $8.7_{-0.9}^{+8.4}$ & \nodata \\
J1425$-$5723 & \dataset[ADS/Sa.CXO#obs/12686]{12686} & 5 &8.9& 53 & $0.21_{-0.02}^{+0.05}$ & 20\\
J1650$-$4341 & \dataset[ADS/Sa.CXO#obs/13777]{13777} & 8 &3.9& 124 & $6.4_{-2.1}^{+5.0}$ & \nodata\\
J1653$-$4315 & \dataset[ADS/Sa.CXO#obs/13774]{13774} & 5 &6.3& 98 & $2.5_{-0.8}^{+1.9}$ & 0.4 \\
J1702$-$4217 & \dataset[ADS/Sa.CXO#obs/13776]{13776} & 8.6 &6.4& 126 & $6.8_{-1.2}^{+6.3}$ & \nodata\\
J1739$-$3951 & \dataset[ADS/Sa.CXO#obs/12685]{12685} & 5 &6.3& 46 & $0.12_{-0.01}^{+0.02}$ & 30\\
J1755$-$2725 & \dataset[ADS/Sa.CXO#obs/08717]{8717},
\dataset[ADS/Sa.CXO#obs/08718]{8718} & 4 &12& 79 & $1.1_{-0.2}^{+0.5}$ & 1\\
J1810$-$1820 & \dataset[ADS/Sa.CXO#obs/13775]{13775} & 5 &14& 119 & $5.5_{-1.8}^{+4.0}$ & \nodata\\
J1819$-$1510 & 0406450201, 0505240101 & 71.5 & 1.4 & 117 &
 $5.1_{-1.6}^{+3.7}$ & 0.2\\
B1952+29 & 12684 & 5 &6.3& 45 & $0.107_{-0.002}^{+0.005}$ & 30 
\enddata
\tablenotetext{a}{All are \cxo\ observations except for PSR J1819$-$1510,
which was observed with \xmm\ MOS.}
\tablenotetext{b}{The count rate limits are at a 2$\sigma$ (i.e., 95\%)
confidence level in the 0.5--7\,keV energy range.}
\tablenotetext{c}{The surface temperature ($kT_{\rm max}$) and bolometric
luminosity ($L_{\rm max}^{\rm bol}$) upper limits are inferred from the count
rate limits, assuming uniform blackbody emission with an observed radius of
$R_\infty$=14.5\,km and the DM distance in
Table~\ref{tab:candid}. See the text
for the details.}
\tablenotetext{d}{The age lower limits are inferred from the bolometric
luminosity using the minimal neutron star cooling scenario (see
Figure~\ref{fig:cooling}).}
\end{deluxetable*}

Our sample consists of 12 candidates. They are plotted in the \ppdot\ diagram
in Figure~\ref{fig:ppdot} and their properties are listed in
Table~\ref{tab:candid}. While the pulsars have large characteristic ages of
$\sim10^8$\,years, we note that their true age could be much younger,
particularly if they were born spinning slowly \citep[see][]{nrb+07,bnk14}.
There is an archival \cxo\ on-axis ACIS-S observation of PSR J1355$-$6206 with
3\,ks exposure, and PSR J1755$-$2725 was located far off-axis in two \cxo\
ACIS-I exposures with a total of 4\,ks. PSR J1819$-$1510 falls onto the edge
of the field of view in two \xmm\ MOS exposures. After filtering the
background flaring periods and correcting for vignetting, we obtained an
effective total exposure of 72\,ks for this source. The \xmm\ PN data were not
used in this study, as the total exposure is only 21\,ks. For the remaining
nine candidates, we obtained \cxo\ ACIS-S observations with 5\,ks exposure
each, except for PSRs J1702$-$4217 and J1650$-$4341, which have roughly 8\,ks
each. The observation IDs (ObsIDs) and effective exposures are listed in
Table~\ref{tab:results}. Note that observations of three sources: PSRs
J0609+2130, J1355$-$6206, and B1952+29 were recently reported in an
independent study \citep{gha+13}.

We carried out standard \cxo\ data reduction with CIAO v4.5 and CALDB v4.5.8.
We reprocessed the data using the task \texttt{chandra\_repro} and checked
that there was no background flaring occuring during the \cxo\ observations;
all data are therefore used in the analysis. The \xmm\ data reduction was
performed with SAS v13.5\footnote{\emph{XMM-Newton} SAS is developed and
maintained by the Science Operations Centre at the European Space Astronomy
Centre and the Survey Science Centre at the University of Leicester.}. We used
the task \texttt{emchain} to reprocess the MOS data, then applied filtering
using standard flags and rejected periods with high background. The analysis
is restricted to events with PATTERN $\leq12$. Finally, we generated X-ray
images in the 0.5--7\,keV energy range for source detection. 

\section{Results and Discussion}
\label{sect:discuss}
The X-ray images show no obvious X-ray sources at the radio pulsar positions.
To establish the pulsar flux limits, we extracted events from 5\arcsec\ and
70\arcsec\ radius source and background regions, respectively, for the 10
on-axis \cxo\ observations. The 5\arcsec\ radius was chosen to ensure that the
region encircles over 95\% of the flux. A larger source region of 10\arcsec\
radius was used for PSR J1755$-$2725, due to the large off-axis angle. For the
\xmm\ observations of PSR J1819$-$1510, we used a 60\arcsec\ radius source
region and excluded a background source inside. We follow \citet{bnk14} to
compute the detection limits at a 2$\sigma$ confidence level, using the task
\texttt{aprates} in CIAO, which is based on Bayesian statistics. The results
are reported in Table~\ref{tab:results}.

\begin{figure*}[!ht]
\epsscale{1.1}
\plottwo{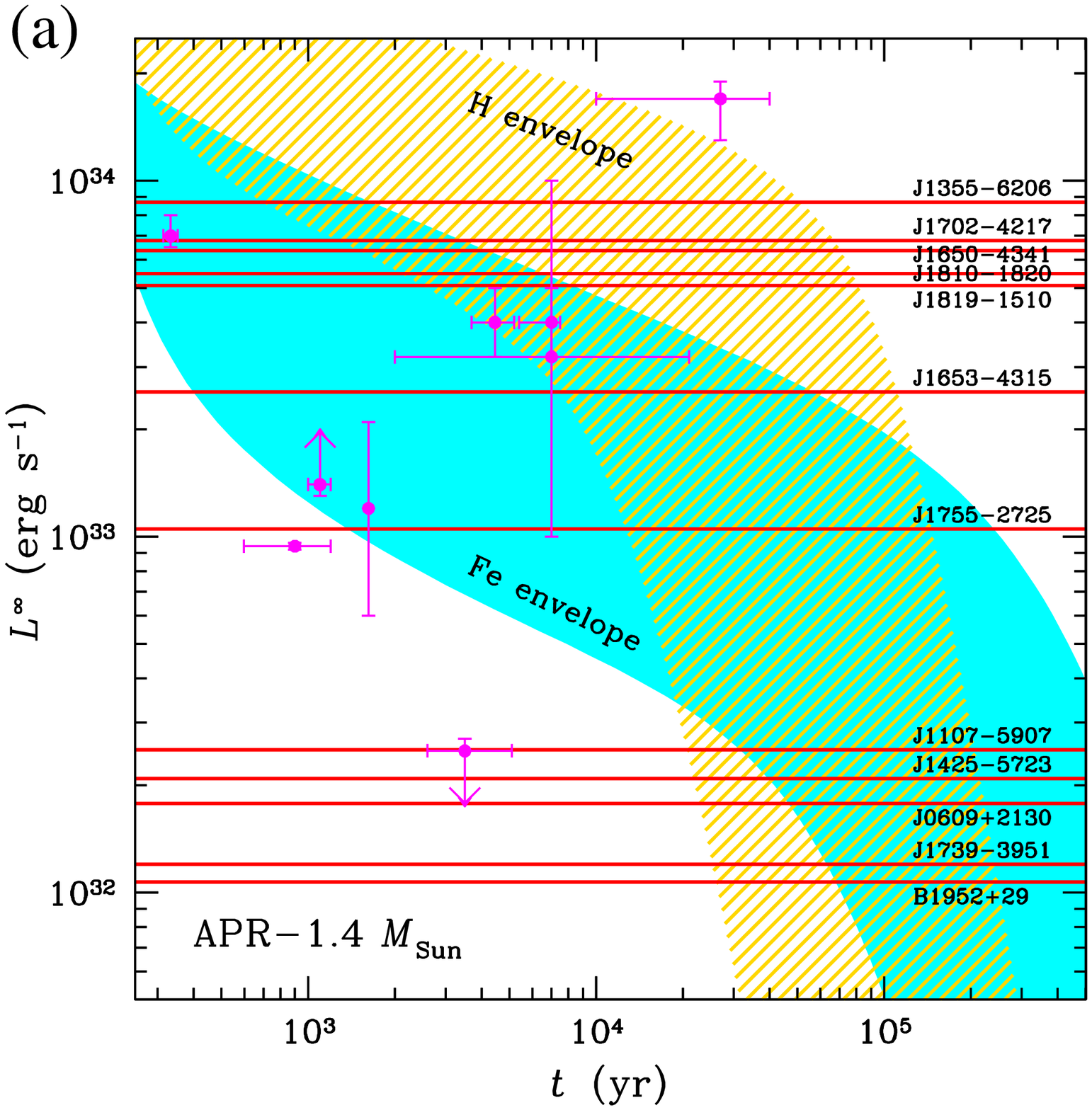}{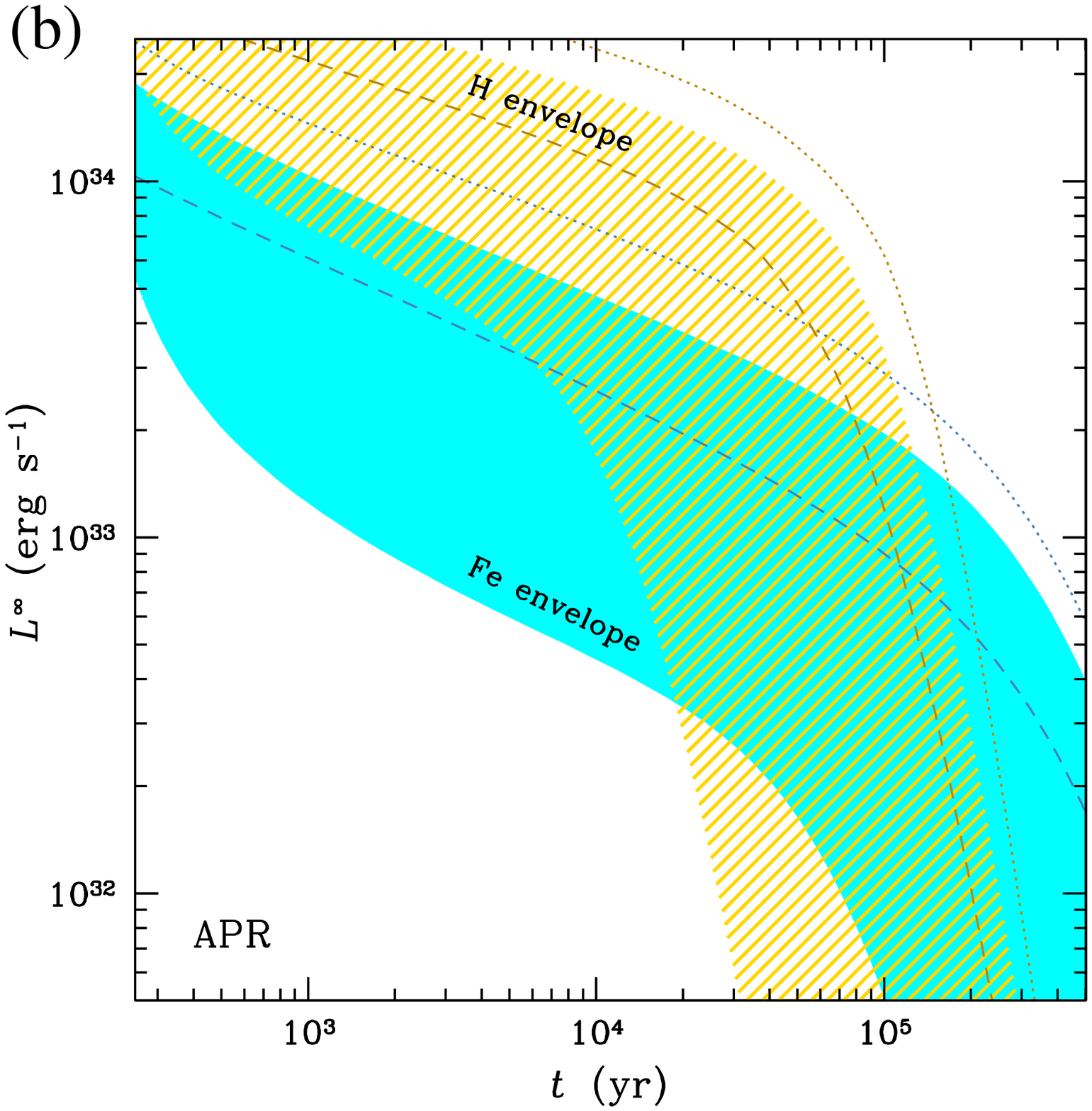}
\caption{\footnotesize (a): Evolution of neutron star bolometric luminosity in
the standard cooling scenario with 1.4\,$M_\odot$, $R=11.6$\,km (i.e.,
$R_\infty$=14.5\,km for a redshift of 1.25), and hydrogen and iron envelopes
(see the text for the details). For each region, the upper bound is obtained
by only considering superconducting protons and the lower bound by considering
superfluid neutrons as well. The red lines indicate the luminosity limits of
the orphaned CCO candidates listed in Table~\ref{tab:results}. The magenta
points show the age and luminosity values of confirmed CCOs, from high to low
luminosity: XMMU~J173203.3$-$344518 in HESS~J1731$-$347 \citep{ksp+15};
CXOU~J232327.9+584842 in Cas~A \citep{fhm+06,hh10}; CXOU~J185238.6+004020 in
Kes~79 \citep{sss+04,vrp+13}; RX~J0822.0$-$4300 in Puppis~A
\citep{bpw+12,vrp+13}; 1E~1207.4$-$5209 in PKS 1209$-$51/52
\citep{rmk+88,vrp+13}; CXOU~J160103.1$-$513353 in G330.2+1.0
\citep{mgd+01,pkp+09}; 1WGA~J1713.4$-$3949 in G347.3$-$0.5
\citep{lsg+03,cdb+04,fkp+12}; XMMU~J172054.5$-$372652 in G350.1$-$0.3
\citep{gts+08,lsg+11}; and CXOU~J085201.4$-$461753 in G266.1$-$1.2
\citep{kps+02,acd+15}.
(b): Cooling curves of 1.2\,$M_\odot$ (dotted lines) and 1.9\,$M_\odot$ (dashed
lines) neutron stars calculated assuming superconducting protons only.
\label{fig:cooling}}
\end{figure*}

The count rate limits above can be converted to luminosity limits using
a spectral model. Following \citet{gha+13}, we assume uniform blackbody
emission from the entire star with an observed radius of $R_\infty=14.5$\,km
(which corresponds to a neutron star radius of 11.6\,km for a gravitational
redshift of 1.25). The pulsar dispersion measure (DM) distance is adopted and
the X-ray absorption column density (\nh) is inferred from DM using the
empirical relation $N_\mathrm{H}\rm\;(10^{20}\,cm^{-2})=0.30^{+0.13}_{-0.09}
\;DM \;(pc\,cm^{-3})$ \citep{hnk13}. We generated spectral response files for
individual observations and determined the flux limits in the \texttt{Sherpa}
environment \citep{fds01} by comparing the expected count rates with the
observed limits. The surface temperature and bolometric luminosity limits are
listed in Table~\ref{tab:results}. The detection limits of PSRs J0609+2130,
J1355$-$6206, and B1952+29 are compatible with those reported in the previous
study \citep{gha+13}. As shown in the table, the luminosity limits are
$\sim10^{32}$\ergs\ for nearby ($<2$\,kpc) sources and a few times
$10^{33}$\ergs\ for the rest. In Figure~\ref{fig:cooling}, these results are
compared with the observed bolometric luminosities of known CCOs. The latter
are generally above $10^{33}$\ergs, suggesting that our candidates are
unlikely to be young CCOs.

As mentioned, the characteristic age is not a good indicator of the true age
of a pulsar. We therefore attempt to constrain the age by comparing the
luminosity limits above with standard neutron star cooling curves. We employed
a cooling code based on \citet{gyp01} with some modifications as described in
\citet{heh+15}. The temperature evolution of an isolated neutron star is
determined by the relativistic equations of energy balance and heat flux
\citep[see, e.g.,][]{yp04}. Two important factors affecting the evolution are
the composition of the outer layer (i.e., the envelope) and whether the
stellar interior is superfluid and/or superconducting. For the former, the
envelope serves as a heat blanket \citep{gpe82}. If it is composed of light
elements, which have higher thermal conductivity, it would be more heat
transparent \citep{pyc+03}. However, the exact composition of the envelope is
uncertain.
For the second factor, superfluidity and superconductivity suppress heat
capacities and some emission mechanisms, such as modified Urca processes. On
the other hand, emission due to the Cooper pairing of nucleons would be
enhanced when the temperature decreases just below the critical value
\citep[see][for reviews]{yp04,pgw06}.

Here we consider a $1.4\,M_\odot$ and 11.6\,km radius (corresponding to an
observed radius $R_\infty=14.5$\,km) star built with the
Akmal-Pandharipande-Ravenhall (APR) equation of state (EOS) \citep{apr98} and
the standard (also known as minimal) cooling scenario, in which neutrino
emission by fast direct Urca processes does not take place \citep[see,
e.g.,][]{yp04,pgw06}. We show in Figure~\ref{fig:cooling} two shaded regions,
one for a maximally thick $10^{-7}\,M_\odot$ hydrogen envelope and one for an
iron envelope. The upper boundary of each region is determined by a cooling
model with only superconducting protons (gap model CCDK) and the lower
boundary accounts for superfluid neutrons as well (singlet gap model SFB and
triplet gap model TToa), which induce more rapid cooling \citep[see][for
details]{heh+15}. To illustrate the effect of different masses, we plot in the
figure inset cooling curves of $1.2\,M_\odot$ and $1.4\,M_\odot$ stars for
models with only superconducting protons.

The model cooling curves are compared with the candidates' luminosity limits
in Figure~\ref{fig:cooling} to constrain the source age. The result slightly
depends on the neutron star mass in the model, but the main uncertainty comes
from the whether or not neutrons in the core become superfluid. The inferred age
lower limits are listed in Table~\ref{tab:results}. The results are not very
constraining. For nearby objects that have good luminosity limits, we deduce
age limits of $\sim2 \times10^4$\,years, but no useful limits are obtained for
the rest, and deeper X-ray observations are needed for further investigation.
Unlike the CCO candidates in our previous study \citep{bnk14}, none of the
sources here coincide with known SNRs, and we do not find evidence of remnant
emission from the X-ray data. This indeed gives a hint that the sources are
likely older than $\sim10^5$\,years, such that the SNRs have already dissipated
in the interstellar environment. 

The non-detection of X-ray emission from our sample of weak-field radio
pulsars seems to support the idea that the two are not evolutionary connected
classes of neutron stars, similar to what previous studies have suggested
(e.g., \citealt{gha+13}, \citealt{bnk14}). It was proposed that the magnetic
fields of CCOs could be buried by accretion from supernova materials at birth
\citep{hg10,ho11,ho13a}, resulting in the suppression of radio emission. As
the field diffuses back to the surface, the radio emission would subsequently
turn on so that a CCO becomes a radio pulsar. To investigate the effect of
magnetic field diffusion on the neutron star spin evolution, we applied a
model described in \citet{ho11} to solve the induction equation. We assumed
that the $B$-field is confined in the crust and is initially buried at a high
density of $10^{13}$\,g\,cm$^{-3}$, which corresponds to a depth of
$\sim400$\,m or an accreted mass of $\sim5\times 10^{-4}\,M_\odot$
\citep[see][]{ho11}. We consider different initial surface $B$-field strengths
from $10^{12}$\,G to $10^{14}$\,G and the resulting spin evolution is
overplotted in the \ppdot\ diagram in Figure~\ref{fig:ppdot}. We found that
even at this large burial depth, the field emerges with a relatively short
timescale of a few $\times10^4$\,years, resulting in a vertically upward track
in the diagram. More importantly, a CCO with an emerging surface field is
expected to pass through the \ppdot\ parameter space of our candidates (i.e.,
$B\approx 10^{10}$--$10^{11}$\,G) at a young age of the order of $10^4$\,yr,
such that it should remain hot and luminous, with the parent SNR probably
still visible. If this model is true, our candidates are likely to just be old
ordinary pulsars rather than orphaned CCOs. Moreover, we expect that when an
aged CCO turns on as a radio pulsar, it will have a stronger $B$-field than
those of our candidates.

For further study, identifying CCO descendants would require X-ray
observations of more radio pulsars to search for thermal emission \citep[see
also][]{gha+13}. In particular, it would be useful to extend the survey to
include sources with $B\gtrsim 10^{11}$\,G. On the other hand, it is more
difficult to detect an orphan CCO if the radio emission never turns on
\citep[see][]{hbg13}. This is best done with all-sky surveys in X-rays, such
as the upcoming \emph{eROSITA} mission \citep{mpb+12}. Finally, pulsar braking
index measurements from long-term timing could reveal a growing dipole field
\citep[e.g.,][]{elk+11}, which could indicate $B$-field diffusion to the
surface \citep[see][]{ho11}.

\section{Conclusions}
We present an X-ray study of all isolated radio pulsars within 100\,pc of the
Galactic plane, which has spin-down-inferred $B$-fields weaker than
$10^{11}$\,G, with the aim of searching for CCO-like thermal X-ray emission.
None of our 12 candidates were detected in the \cxo\ and \xmm\ observations
and we obtained bolometric luminosity limits of $10^{32}$--$10^{34}$\ergs. The
limits were compared with a standard neutron star cooling model to constrain
the source ages. Together with the lack of known associated SNRs, we conclude
that all candidates are unlikely to be young CCOs. This result supports the
idea that young CCOs and currently known weak-field radio pulsars are not
connected through evolution. This could be a consequence of field burial
by supernova materials. We model the pulsar spin evolution in this scenario
and show that the surface $B$-field could reemerge rapidly with a timescale
shorter than 100\,kyr. It could therefore be fruitful to include radio pulsars
with slightly stronger fields ($B\sim10^{11}$\,G) in future searches for CCO
descendants.

\acknowledgements
We thank the referee for a careful reading and useful comments. 
This work was funded in part by the NASA \emph{Chandra} grant GO2-13079X
awarded through Columbia University and issued by the \emph{Chandra X-ray
Observatory} Center, which is operated by the Smithsonian Astrophysical
Observatory for and on behalf of NASA under contract NAS8-03060. A portion of
the results presented was based on observations obtained with
\emph{XMM-Newton}, an ESA science mission with instruments and contributions
directly funded by ESA Member States and NASA. The scientific results reported
in this article are based in part on observations made by the \emph{Chandra
X-ray Observatory}. This research has made use of the NASA Astrophysics Data
System (ADS) and software provided by the Chandra X-ray Center (CXC) in the
application package CIAO. WCGH acknowledges support from STFC in the UK.
V.M.K.\ receives support from an NSERC Discovery Grant and Accelerator
Supplement, the Centre de Recherche en Astrophysique du Quebec, an R. Howard
Webster Foundation Fellowship from the Canadian Institute for Advanced Study,
the Canada Research Chairs Program, and the Lorne Trottier Chair in
Astrophysics and Cosmology.


\end{document}